\documentclass{agujournal}
\draftfalse

\usepackage{array,multirow}
\usepackage{amsmath}
\usepackage{color,soul}
\journalname{Journal of Geophysical Research: Planets}

\begin{document}

\title{Where does Titan Sand Come From: Insight from Mechanical Properties of Titan Sand Candidates}

\authors{Xinting Yu\affil{1}, Sarah M. H\"orst\affil{1}, Chao He\affil{1}, Patricia McGuiggan\affil{2}, Bryan Crawford\affil{3}}

\affiliation{1}{Department of Earth and Planetary Sciences, Johns Hopkins University, 3400 N. Charles Street, Baltimore, Maryland 21218, USA.}
\affiliation{2}{Department of Materials Science and Engineering, Johns Hopkins University, 3400 N. Charles Street, Baltimore, Maryland 21218, USA.}
\affiliation{3}{Nanomechanics, Inc., 105 Meco Ln, Oak Ridge, Tennessee 37830, USA.}

\correspondingauthor{Xinting Yu}{xyu33@jhu.edu}

\begin{keypoints} 
\item Tholin has a high elastic modulus and hardness, but low brittleness compared to common polymers due to its complex cross-linked structure.
\item With a magnitude lower modulus, hardness and fracture toughness than silicate sand, tholin may be hard to transport long distances on Titan.
\item Under Titan conditions, water ice and simple organics are also mechanically weak and thus may be even more difficult to transport on Titan.
\end{keypoints}

\begin{abstract}
Extensive equatorial linear dunes exist on Titan, but the origin of the sand, which appears to be organic, is unknown. We used nanoindentation to study the mechanical properties of a few Titan sand candidates, several natural sands on Earth, and common materials used in the Titan Wind Tunnel, to understand the mobility of Titan sand. We measured the elastic modulus (E), hardness (H), and fracture toughness ($\mathrm{K_c}$) of these materials. Tholin's elastic modulus (10.4 $\pm$ 0.5 GPa) and hardness (0.53 $\pm$ 0.03 GPa) are both an order of magnitude smaller than silicate sand, and is also smaller than the mechanically weak white gypsum sand. With a magnitude smaller fracture toughness ($\mathrm{K_c=0.036\pm0.007}$ MPa$\cdot$$\mathrm{m^{1/2}}$), tholin is also much more brittle than silicate sand. This indicates that Titan sand should be derived close to the equatorial regions where the current dunes are located, because tholin is too soft and brittle to be transported for long distances.

Plain language summary: Sand dunes, which are probably made of organic materials, are observed on Titan in the equatorial region, but the origin of the organic sand is a mystery. We measured mechanical properties of several Titan sand analogs, so that we can estimate their ability to transport on Titan's surface and help us constrain the source region of Titan sand. We found out that most of the possible candidates of Titan sand, including tholin (Titan aerosol analog), water ice, and some simple organics, are all less stiff, softer and more brittle than the silicate sand being transported on Earth's surface. This suggests that sand on Titan may be too weak mechanically to transport long distances on Titan. Thus it is unlikely for Titan sand to originate from the polar regions of Titan, where the methane lakes and seas are located and have been suggested as one possible formation location.
\end{abstract}

\section{Introduction}
Across the Solar System, many planetary worlds have aeolian processes despite the wide variety of environmental conditions present on these bodies. These bodies include: Venus, Earth, Mars (Greeley \& Iversen, 1985), Titan (Lorenz et al., 2006), Neptune's moon Triton (Smith et al., 1989), Pluto (Telfer et al., 2018), and possibly comet 67P/Churyumov-Gerasimenko (Thomas et al., 2015). Other than the environmental conditions, the aeolian processes on icy bodies (Titan, Triton, and Pluto) differ from those on terrestrial bodies (Venus, Earth, and Mars) because of the differences in the dune-forming materials. On terrestrial bodies, the materials that get transported are mainly silicate sand (weathering and erosion products of silicate rocks), while the materials that are transported on icy bodies could be different. For example, the wind streaks on Triton are possibly composed of dark complex hydrocarbons (Smith et al., 1989), the dunes on Pluto are made of methane ice (Telfer et al., 2018), and the ripples and wind tails on comet 67P could be made of organic-rich materials associated with opaque minerals on the surface (Capaccioni et al., 2015). On Titan, the dune-forming materials are most likely organics produced by photochemistry in the atmosphere (Soderblom et al., 2007; H\"orst, 2017) or abundant ices that form the crust of Titan. Silicate sand is known to have high resistance to abrasion due to its hardness (Mohs hardness around 6 to 7), which might be the reason that it can be transported for long distances without being abraded to dust (Bagnold, 1941). However for icy bodies like Titan, we do not know the basic mechanical properties of the organic sand or ice on the surface, so we cannot infer its transport capabilities.

Titan's sand particle sizes are first estimated to be around 100--300~$\upmu$m based on calculation of the optimum particle diameter range that results the minimum threshold wind speed on Titan (Lorenz et al., 2006). Lorenz (2014) suggests that plausibly decreased particle density or increased cohesion between particles could lead to higher optimum diameter up to around 500--600~$\upmu$m; Yu et al., (2017a) did find the cohesion forces of tholin larger than those of silicate sand and materials used in the Titan Wind Tunnel. Burr et al., (2015) modified the threshold friction speed function using experimental results in the Titan Wind Tunnel and they found an optimum diameter around 200--300~$\upmu$m. All those previous studies indicate that the size of the Titan sand particles should be on the order of hundreds of microns. So it is a puzzle how the small aerosol particles produced in Titan's atmosphere (up to 1~$\upmu$m, Tomasko et al., 2005) are transformed into these large, sand-sized particles on Titan's surface (Soderblom et al., 2007). Barnes et al. (2015) proposed four mechanisms for the transformation: sintering, lithification and erosion, flocculation, and evaporation. The sintering and lithification and erosion mechanisms could happen in subaerial conditions while the flocculation and evaporation need subaqueous environments. However, current lakes and seas on Titan are mainly at high latitudes while the longitudinal dunes are thousands of kilometers away in the equatorial region. Thus, if sand particles on Titan were produced in the current lakes and seas by subaqueous mechanisms, they need to be mechanically strong enough to travel long distances to the equator. 

Therefore, it is important to quantify the mechanical behaviors of Titan sand analog materials so that we can better understand the origin of Titan sand particles and their transportation capacities. Laboratory-produced Titan aerosol analogs (so-called ``tholins'') could be compositionally similar to Titan sand (Barnes et al., 2015; Yu et al., 2017a), but are usually produced in low yields (Cable et al., 2012) and thus are difficult to quantify mechanically using macroscopic approaches. This makes nanoindentation an ideal method to quantify the mechanical behaviors of the thin tholin films. Nanoindentation is a technique that uses small loads (on the order of mN) and small tip size (tip radius on the order of 100 nm), resulting in a nanometer scale indentation area, and is used widely for quantifying mechanical properties of small volumes of materials.

Evaporites are also possible candidate materials for Titan sand (Barnes et al., 2015). Titan's evaporites may be made of acetylene, ethylene or butane (Cordier et al., 2013, 2016; Singh et al., 2017), however, these materials are not stable solids under room temperatures on Earth, and their solid mechanical properties also have not been measured under low temperatures.

Another possible candidate for Titan sand is water ice. Even though the dune-making materials on Titan appear to be dominated by a spectrally ``dark brown'' organic unit in Cassini's Visual and Infrared Mapping Spectrometer (VIMS) data, with little water ice ``dark blue'' spectral signature (Soderblom et al., 2007; Barnes et al., 2008), it cannot be ruled out that the individual sand particles are water ice grains coated with a thin layer of organics because the infrared penetration depth is at most tens of microns (Barnes et al., 2008).

Kuenen (1960) found that various mechanical properties are involved in mechanical abrasion in aeolian and aqueous transport. For relatively soft materials, the dominant abrasion mechanism is ``grinding'' (where hardness of the material dominates); in this case, a change of grain size or wind speed would not substantially affect the abrasion rate. While for relatively hard materials like quartz, its brittleness makes ``chipping'' (or ``spalling'') the dominate mechanical abrasion mechanism under aeolian transport (when impacts dominate over direct fluid drag). ``Chipping'' of quartz grains slows down with increasing roundness, decreasing grain size, and decreasing wind speeds. While in aqueous transport, where impacts are minimal, quartz erodes very slowly because of its high hardness. Thus it is important to characterize both the mechanical hardness and brittleness of Titan sand analogs, so that we can better assess aeolian versus fluvial transportation on Titan.

The materials and nanoindentation methods are described in Section 2.1--2.2. In Section 3.1, we compare the measured elastic modulus and hardness of various materials. Measured fracture toughness of selected materials are compared in Section 3.2. We discuss the extrapolation of mechanical properties of tholin from room temperature to Titan's surface temperature in Section 4.1. Finally, we list all the possible candidate materials for Titan sand and discuss their capability of transportation on Titan in Section 4.2. 

\section{Methods}
\subsection{Materials and Preparation}
We used a variety of analog materials, both lab-created and naturally found, in order to simulate materials being transported on both terrestrial and icy bodies. We used a few natural sands on Earth, including silicate beach sand, carbonate sand, and white gypsum sand as terrestrial sand analogs.

For Titan, we used both the laboratory produced tholin (He et al., 2017) and some simple solid organic materials as analog materials. Tholin was produced using the Planetary HAZE Research (PHAZER) experimental system at Johns Hopkins University, with a 5\% $\mathrm{CH_4/N_2}$ cold gas mixture (around 100 K) in a glow plasma discharge chamber (pressure: 3 Torr, flow rate of gas mixture: 10 sccm, He et al., 2017). The produced tholin simulates the aerosol on Titan and is mixture of complex organic compounds.

Different types of simple organics were used to simulate simple atmospheric condensates and /or evaporites on Titan. Titan's evaporites could be made of acetylene, ethylene, or butane (e.g. Cordier et al., 2013). Since we are measuring the materials under room temperature, we chose the following simple organics since they are stable solids under room temperature. Some of the following organics may exist in Titan's atmosphere and some are identified in tholin samples before. Three polycyclic aromatic hydrocarbons (PAHs) made of different numbers of fused benzene rings are used: naphthalene (two rings, possibly present in Titan's upper atmosphere, Waite et al., 2007), phenanthrene (three rings, its mass peak, $\sim$170 amu, possibly present in Cassini Plasma Spectrometer data, Waite et al., 2007), and coronene (six rings). One polyphenyl, biphenyl, made of two non-fused benzene rings, was also used to compare with naphthalene (two fused benzene rings). Two nitrogen-containing organics, adenine and melamine, both of which have been identified in tholin samples from a different experimental setup (H\"orst et al., 2012, He \& Smith, 2013, 2014a, b) were used to test the effect of nitrogen inclusion on the mechanical properties of simple organics.

We also investigated analog materials used in planetary wind tunnels (Yu et al., 2017b), including chromite, basalt, quartz sand, glass beads, gas chromatograph packing materials (called GC), activated charcoal, instant coffee, walnut shells, and iced tea powder. 

All the above materials are also summarized in Table \ref{table:materials}.

Tholin was deposited as a thin homogenous film on mica discs (10 mm diameter). The film is very smooth (RMS roughness is $\sim$1 nm, measured by Atomic Force Microscopy, AFM) and has a thickness of approximately 1.3 $\mu$m. We also collected tholin particles from the chamber wall in a dry $\mathrm{N_2}$ glove box ($\mathrm{O_2\textless1\ ppm,\ H_2O\textless \ 1ppm}$).

The laboratory-produced tholin film is used directly for the measurements. In contrast, the collected tholin particles and other material particles (Table \ref{table:materials}) needed to be mounted and polished before measurement. The procedure for preparing the particles is as follows: the particles were embedded in an epoxy matrix using a vacuum mounting system in cylinder sample stubs (1.25" diameter). The samples were cured in ambient atmosphere overnight, resulting in a composite of particles in a hardened epoxy matrix. The samples were then polished to obtain a smooth surface for nanoindentation. For water insoluble materials, the samples were polished using a Tegramin-20 Sample Polisher. The finest polishing size was 40 nm using non-drying colloidal silica suspension. For water soluble materials, we used hand polishing; the finest grain size was 3.5 $\mu$m with 7000 grit silicon carbide paper.

\begin{table}
\caption{Summary of materials used in this study. Basalt is acquired from Pisgah crater and only two major compositions are shown in the table marked with *. Its detailed composition can be found in Friedman (1966). GC indicates gas chromatograph packing materials. GC pink is diatomite, while GC tan is calcined diatomite, it has a different color compared to GC pink (see also Burr et al., 2015; Yu et al., 2017a).}
 \label{table:materials}
 \centering
 \begin{tabular}{c c c}
 \hline
Material Category & Material Name & \ \ \ Structure\\
\hline
\multirow{7}{*}{Titan sand analogs} & Tholin & $\ \ \ \mathrm{C_xH_yN_z}$ \\\\[-1em] \cline{2-3}
\\[-1em] & Naphthalene ($\mathrm{C_{10}H_{8}}$) & \parbox[c]{2.8em}{\includegraphics[height=0.3in]{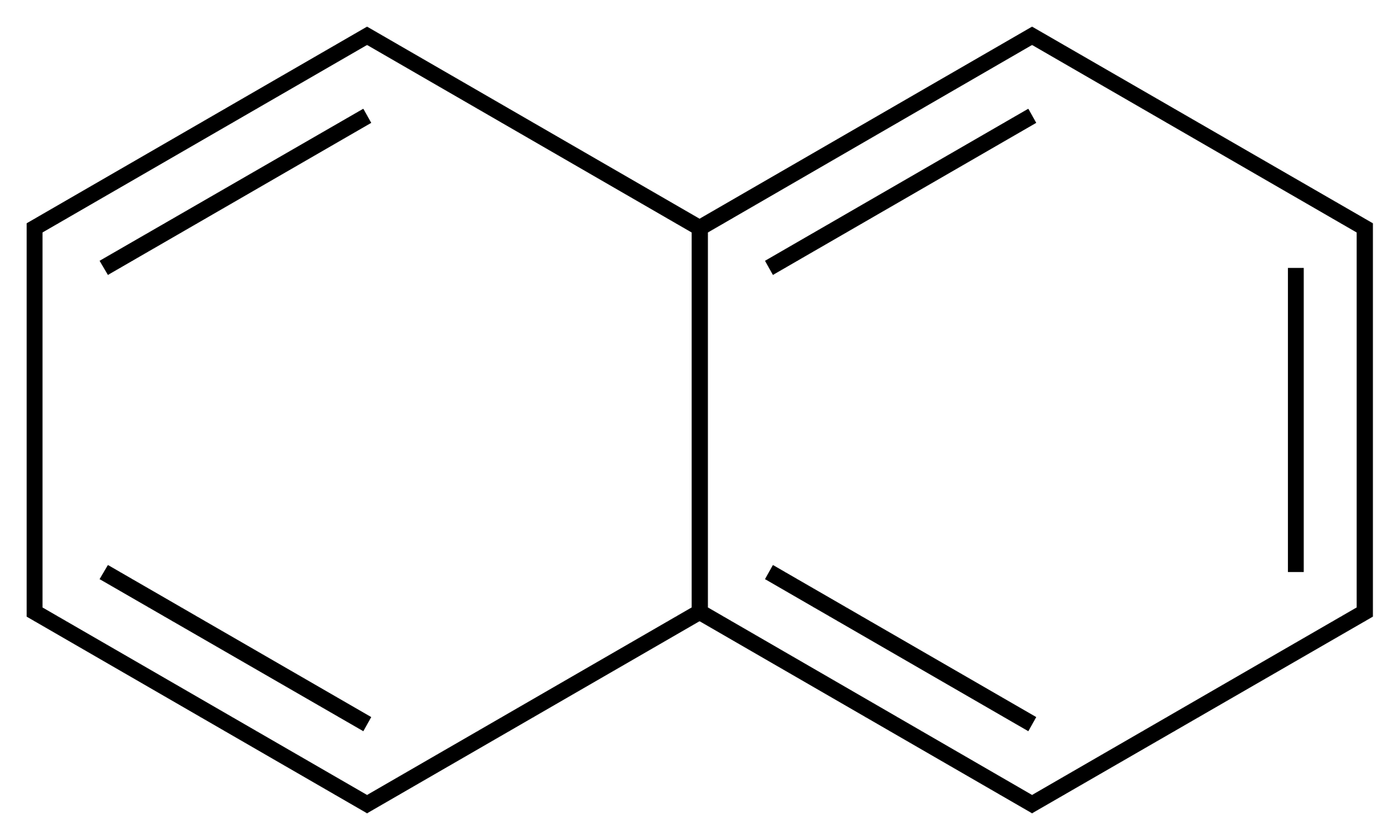}}\\ \\[-1em] \cline{2-3}\\[-1em]
 & Biphenyl ($\mathrm{C_{12}H_{10}}$) & \parbox[c]{4.3em}{\includegraphics[height=0.3in]{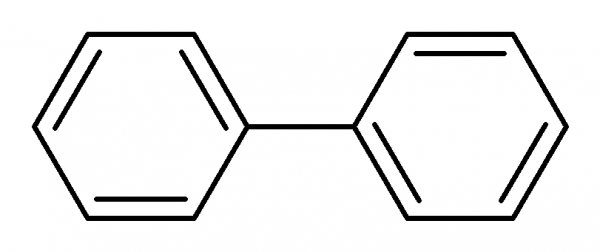}}\ \\\cline{2-3}\\[-1em]
 & Phenanthrene ($\mathrm{C_{14}H_{10}}$) & \parbox[c]{4.3em}{\includegraphics[height=0.4in]{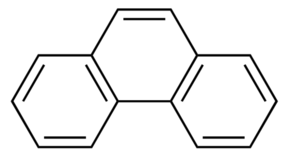}} \\ \cline{2-3}\\[-1em]
 & Coronene ($\mathrm{C_{24}H_{12}}$) & \parbox[c]{3.7em}{\includegraphics[height=0.6in]{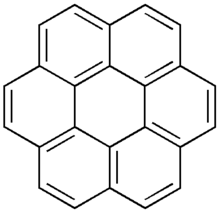}} \\\\[-1em]\cline{2-3}\\[-1em]
 & Adenine ($\mathrm{C_{5}H_{5}N_{5}}$) & \parbox[c]{3.2em}{\includegraphics[height=0.6in]{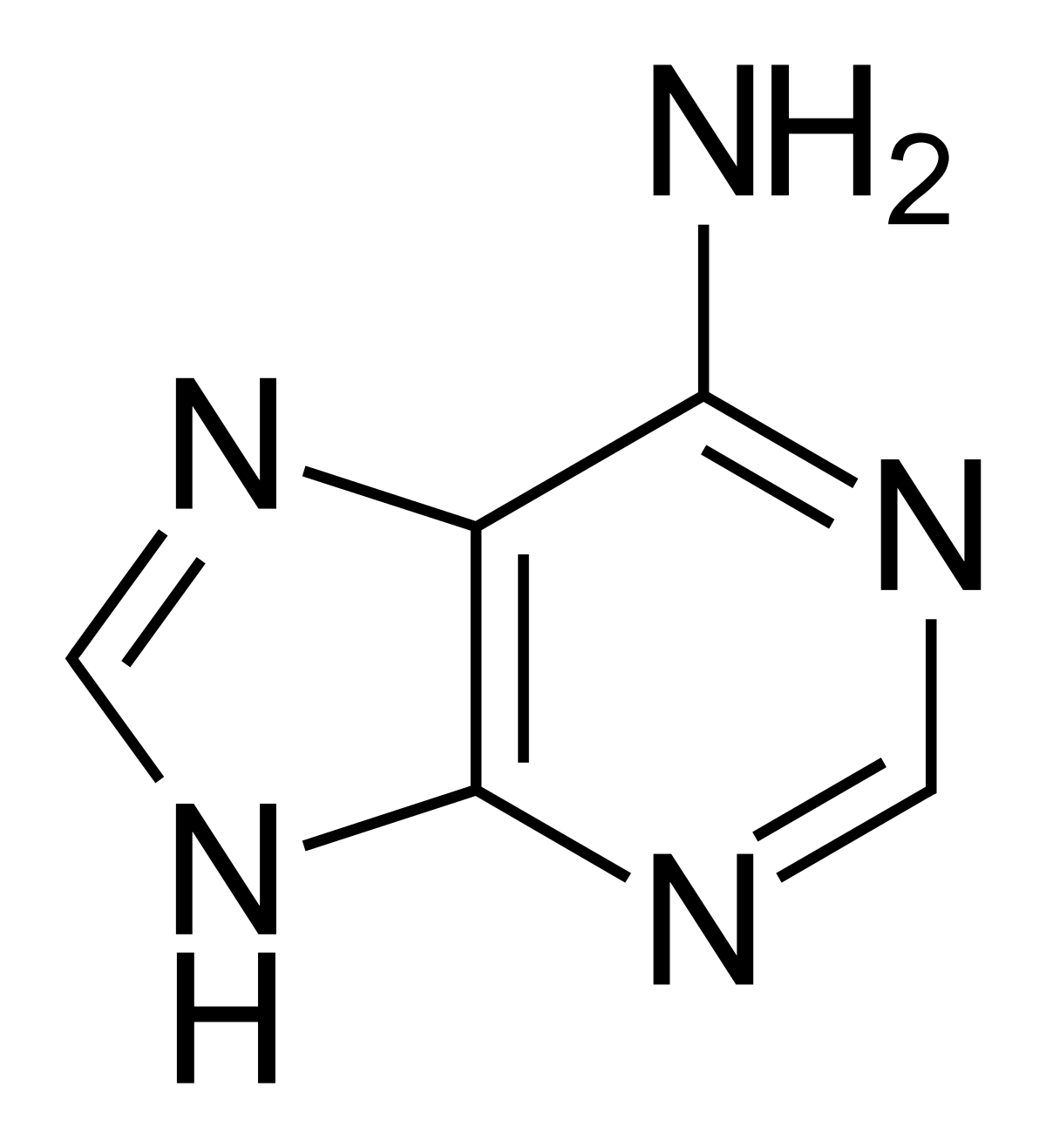}}\\\cline{2-3}\\[-1em]
 & Melamine ($\mathrm{C_{3}H_{6}N_{6}}$) & \parbox[c]{5.4em}{\includegraphics[height=0.6in]{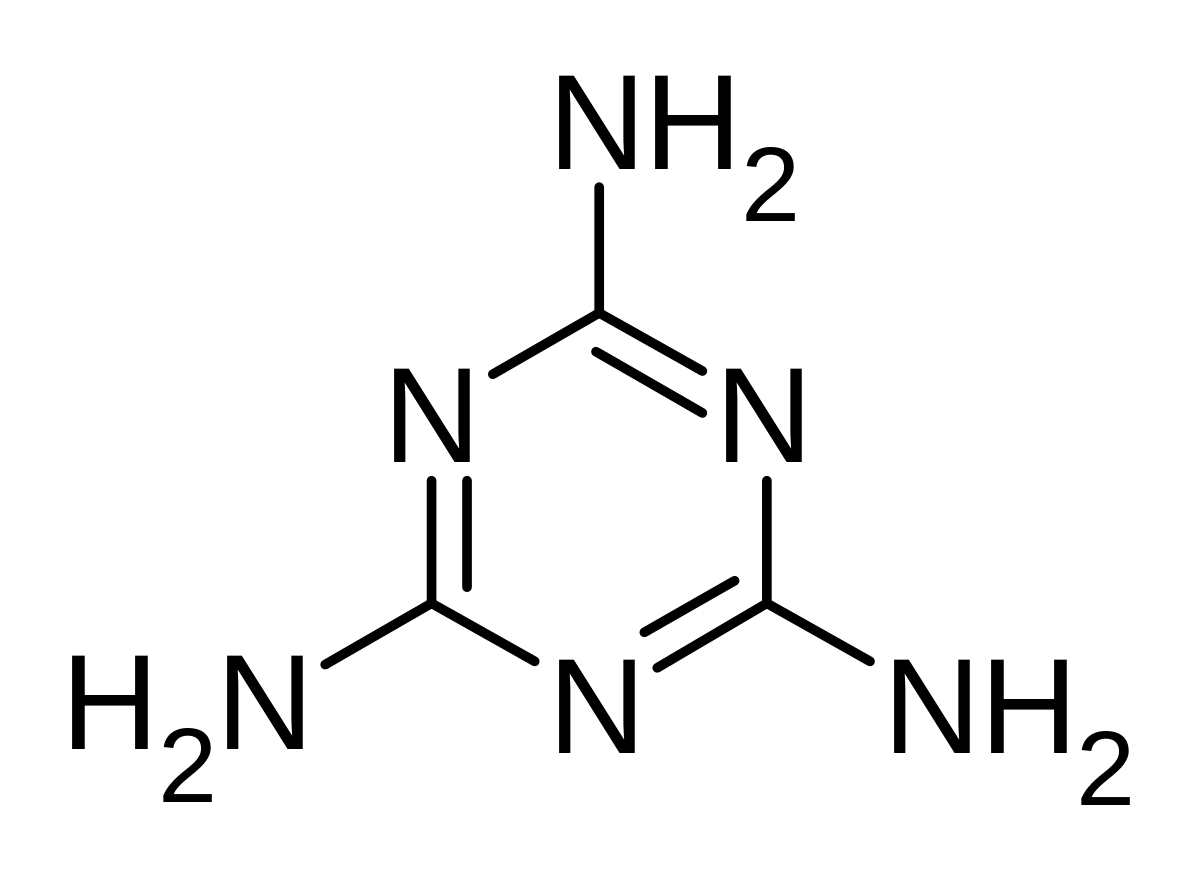}}\\
\hline
\multirow{3}{*}{Natural sand} & Silicate beach sand & \ \ \ mainly $\mathrm{SiO_2}$\\
& Carbonate sand & \ \ \ mainly $\mathrm{CaCO_3}$\\
 & White gypsum sand & \ \ \ mainly $\mathrm{CaSO_4\cdot2H_2O}$\\
 \hline
 & Chromite & \ \ \ mainly $\mathrm{(Fe, Mg, Al)Cr_2O_4}$\\
 & Basalt & mainly $\mathrm{{SiO_2,\ Al_2O_3}^*}$\\
Materials used in & Quartz sand & \ \ \ mainly $\mathrm{SiO_2}$\\
planetary wind tunnels & Glass beads & \ \ \ mainly $\mathrm{SiO_2}$\\
(e.g., Titan Wind & GC pink & \multirow{2}{*}{\ \ \ mainly modified $\mathrm{SiO_2}$}\\
Tunnel, TWT; Martian  & GC tan\\
Surface Wind Tunnel, & Activated charcoal & mainly C\\
MARSWIT, etc.)  & Instant coffee & n/a\\
 & Walnut shells & n/a\\
 & Iced tea powder & n/a\\
 \hline
 \end{tabular}
 \end{table}

\subsection{Nanoindenter and Tips}
We used an iNano Nanoindenter (Nanomechanics, Inc.) for the elastic modulus, hardness, and fracture toughness measurements. The instrument has a maximum load of 50 mN, with a load resolution of 3 nN and a displacement resolution of 0.02 nm. A grid of points on the material were indented and each time the instrument recorded a load-displacement curve. During each load--displacement cycle, the applied load will stop increasing when the maximum load or the maximum penetration depth is reached. We performed all the measurements under room temperature, and then estimated the result for tholin under Titan's surface temperature (94 K).

 \begin{figure}[h]
 \centering
 \includegraphics[width=32pc]{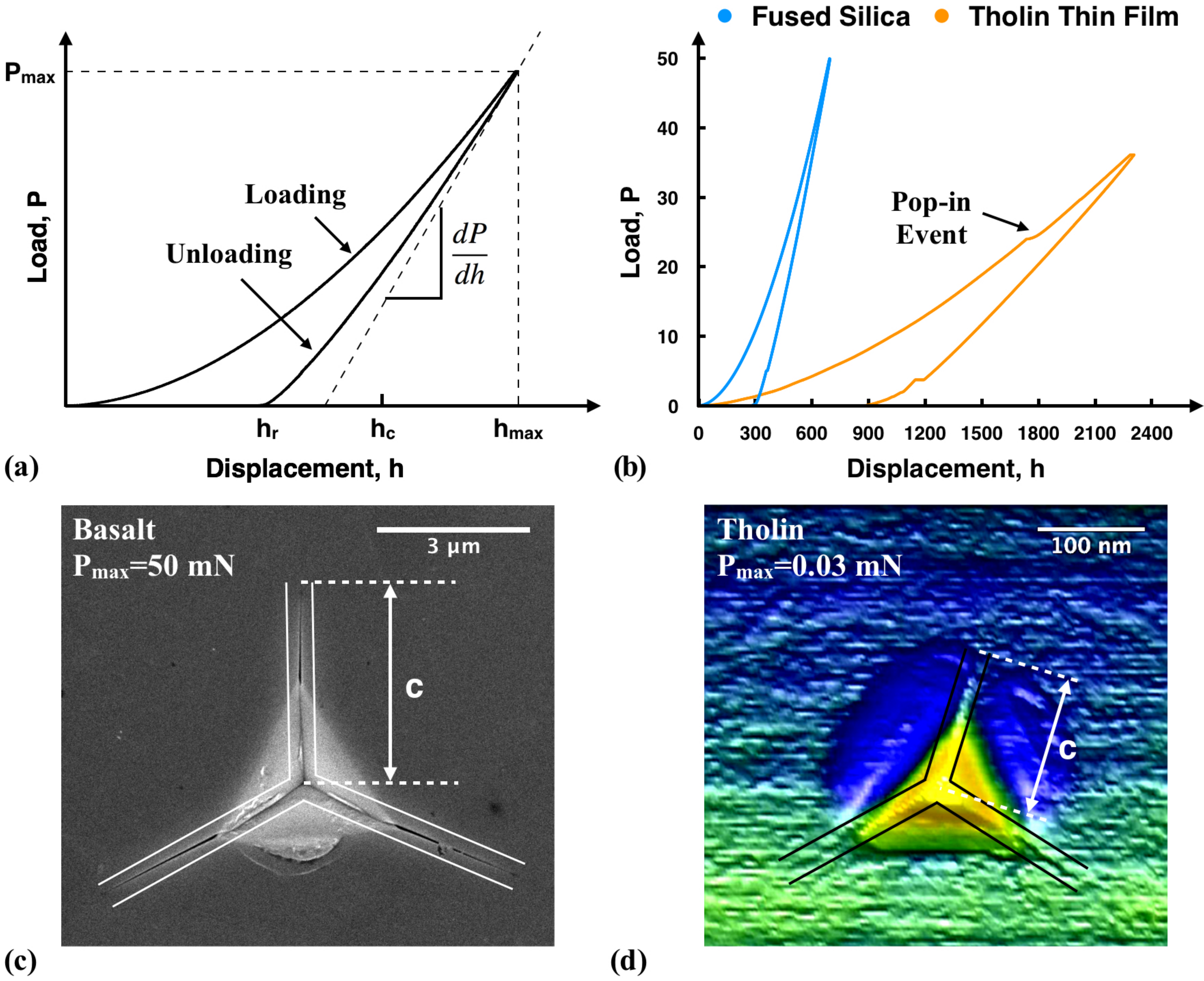}
 \caption{(a) A schematic representation of load (P)--indenter displacement (h) curve in a nanoindentation experiment, where $\mathrm{P_{max}}$ is the maximum load, $\mathrm{h_{max}}$ is the maximum displacement at peak load, $\mathrm{h_{c}}$ is the depth of contact at peak load, and $\mathrm{h_{r}}$ is the residual depth of contact impression after unloading. (b) A comparison between the load--displacement curves of fused silica and tholin thin film with a Berkovich indenter. In the load--displacement curve of tholin, a ``pop-in'' event occurs during loading indicating a fracture event. (c) An SEM image showing cracks generated on a basalt grain after nanoindentation using a cube-corner tip. The maximum load is 50 mN and the crack length is denoted as c. (d) An AFM topographic image showing cracks generated on a thin tholin film with a maximum load of 0.03 mN using a cube-corner tip, the crack length is denoted as c.}
 \label{fig:load_displacement_curve}
 \end{figure}
 
We used a three-sided pyramidal-shaped Berkovich tip, made of single crystal diamond (Micro Star Technologies), for measuring the hardness (H) and Young's modulus (E) of the materials. A schematic representation of a load (P) -- displacement (h) curve for measuring elastic modulus and hardness is shown in Figure \ref{fig:load_displacement_curve}(a). 

The hardness (H) is given by:
\begin{linenomath*}
\begin{equation}
H=\frac{P_{\textrm{max}}}{A},
\end{equation}
\end{linenomath*}
where $\mathrm{P_{max}}$ is the peak load and A is the projected area of contact at peak load. For a Berkovich tip, A can be written as:
\begin{linenomath*}
\begin{equation}
A=24.56h^2_c,
\end{equation}
\end{linenomath*}
where $\mathrm{h_c}$ is the depth of contact. It is related to the maximum indentation depth $\mathrm{h_{max}}$, the stiffness S, and the maximum load $\mathrm{P_{max}}$:
\begin{linenomath*}
\begin{equation}
h_\textrm{c}=h_{\textrm{max}}-\epsilon\frac{P_{\textrm{max}}}{S},
\end{equation}
\end{linenomath*}
where $\mathrm{\epsilon}$ is the geometric constant for the indenter; for a Berkovich tip $\mathrm{\epsilon=0.72}$ (Oliver \& Pharr, 1992).

The reduced elastic modulus, $\mathrm{E_r}$, is related to the stiffness S, which can be obtained from the unloading portion of the load--displacement curve:
\begin{linenomath*}
\begin{equation}
S=\frac{dP}{dh}=\frac{2}{\sqrt{\pi}}E_r\sqrt{A}.
\end{equation}
\end{linenomath*}
The elastic modulus of the sample ($\mathrm{E_s}$) can then be obtained from the reduced modulus:
\begin{linenomath*}
\begin{equation}
\frac{1}{E_r}=\frac{1-{\nu_s}^2}{E_s}+\frac{1-{\nu_i}^2}{E_i},
\end{equation}
\end{linenomath*}
where $\mathrm{E_i}$ is the elastic modulus of the tip, and $\mathrm{\nu_s}$ and $\mathrm{\nu_i}$ are the Poisson's ratios of the sample and the tip.

To obtain a higher accuracy of elastic modulus and hardness, we performed dynamic indentation (further described in Oliver \& Pharr, 2004) through each load-displacement cycle. A small harmonic oscillation was superimposed on the applied static load so the instrument could continuously measure elastic modulus and hardness as a function of displacement. Here we use the constant-strain-rate (CSR) method by applying a constant loading rate over the applied load, which approximates a constant strain rate of 0.2 $\mathrm{s^{-1}}$.

Before and after a set of measurements for our samples, fused silica, a commonly used nanoindentation reference material (with a Young's modulus of 72 GPa), was tested to calibrate the area function of the tip. The hardness and elastic modulus were calculated by the software using Oliver and Pharr (1992) method based on the calibration data and the load--displacement curves. 

Since we use dynamic indentation to obtain elastic modulus and hardness continuously as a function of displacement, the actual modulus and hardness value were taken as the average values over a certain indentation depth range. This depth range depends on the surface effect and the effect of the substrate. Generally the indentation average depth is greater than 50 nm to eliminate the effect on the topmost surface caused predominately by surface roughness. For bulk materials, hardness and elastic modulus are independent of indentation depth so the average depth can be taken for any depth range larger than 50--100 nm. For the tholin thin films (thickness around 1.3 $\mu$m) deposited on a hard mica substrate, the substrate effect starts to show up over an indentation depth of 15\% of the film thickness (e.g., Hay \& Crawford, 2011), where the modulus and hardness values start to increase with increasing indentation depth. Here the depth range for reporting the modulus and hardness values is selected to optimize the material response of interest and minimize the surface and substrate effects (for tholin, the average depth range is 100--150 nm).

We used a much sharper cube corner tip, which has a higher aspect ratio than a Berkovich shaped tip, for measuring fracture toughness. With a half angle of only 35.3$^\circ$ (compared to 65.3$^\circ$ for a Berkovich tip), the use of a cube-corner tip can significantly reduce the cracking threshold of brittle materials (Harding et al., 1995). It is also made of single crystal diamond (Micro Star Technologies). When brittle materials are indented with the sharp cube-corner tip, radial cracks are generated. We varied the maximum indentation loads from 0.03 to 50 mN to measure the fracture toughness of our samples. After the indentations, we used a Bruker Dimension 3100 AFM or an SEM (scanning electron microscopy, JSM-6700F, JOEL Ltd.) to image the indentation and the associated cracks. An example is shown in Figure \ref{fig:load_displacement_curve}(c). The fracture toughness calculations were developed by Lawn et al., (1980) and Anstis et al., (1981):
\begin{linenomath*}
\begin{equation}
K=\alpha (\frac{E}{H})^{0.5}(\frac{P_{\textrm{max}}}{c^{3/2}}),
\end{equation}
\end{linenomath*}
where $\alpha$ is an empirical constant that depends on the geometry of the tip; for a cube-corner tip, $\mathrm{\alpha=0.036}$ (Harding et al., 1995). The crack length, c, can be determined by microscopic imaging methods. The fracture toughness test of the reference material, fused silica, was measured to be 0.58$\pm$0.09 MPa$\cdot$$\mathrm{m^{1/2}}$, which is consistent with the literature value (Harding et al., 1995).

When measuring fracture toughness on tholin thin film deposited on mica substrate, if the indentation depth is over 10\% of the film thickness, the elastic-plastic deformation zone and crack growth may extend to the substrate and affect the accuracy of measurements (e.g., Krabbe et al., 2014). Thus we indented the film at a load of only 0.03 mN, which results in a maximum indentation depth of $\sim$100 nm, smaller than 10\% of the film thickness ($\sim$130 nm). The resulting indentation and cracks are shown in Figure \ref{fig:load_displacement_curve}(d).

\section{Results}
\subsection{Elastic Modulus and Nanoindentation Hardness}
\begin{figure}[h]
 \centering
 \includegraphics[width=37pc]{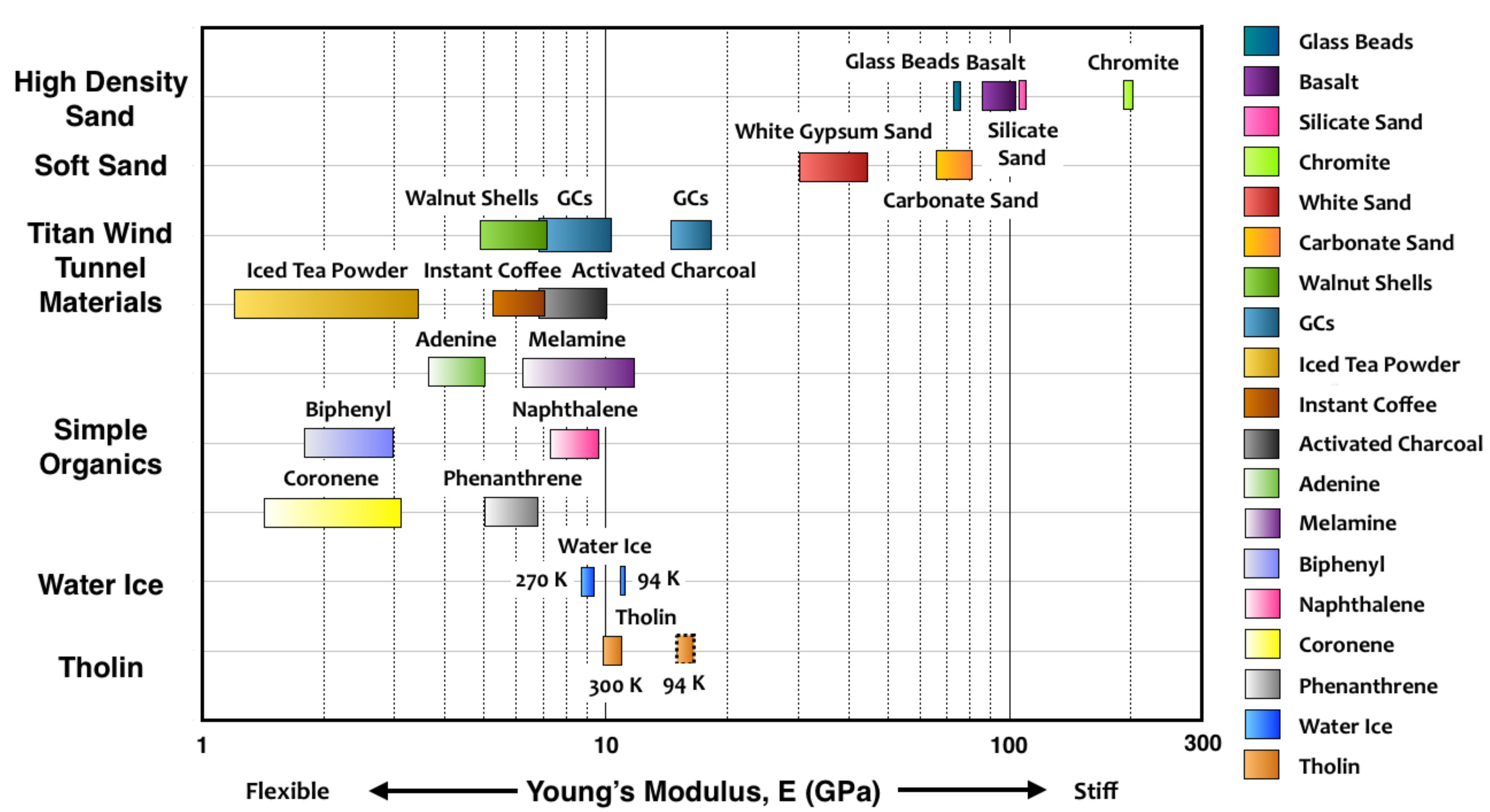}
  \caption{Young's modulus plot for all tested materials. The color bar includes the standard deviation from measurements for each material. Here silicate sand includes both quartz sand (a material used in wind tunnel) and natural silicate beach sand. GCs, the gas chromatography packing materials, include both GC pink and GC tan. Each GC is probably a mixture of two substances, so they each have two sets of characteristic elastic modulus values. Materials are grouped into seven categories: 1) high density materials including glass beads, basalt, silicate sand and chromite in the topmost row; 2) white gypsum sand and carbonate sand in the top second row; 3) low density wind tunnel materials, walnut shells, GCs, iced tea powder, instant coffee, and activated charcoal in the third and fourth rows; 4) nitrogen-containing organics, adenine and melamine, in the fifth row; 5) PAHs (napthalene, phenanthrene, and coronene) and the polyphenol (biphenyl) in the sixth and seventh row; 6) water ice in the eighth row, its elastic modulus under 94 K and 270 K was from the polynomial fitting in Proctor (1966); and 7) tholin in the lowest row, its elastic modulus value under ambient environment (300 K) is measured here and the value under 94 K is extrapolated in Section 4.1. }
 \label{fig:modulus}
 \end{figure}

\begin{figure}[h]
 \centering
 \includegraphics[width=37pc]{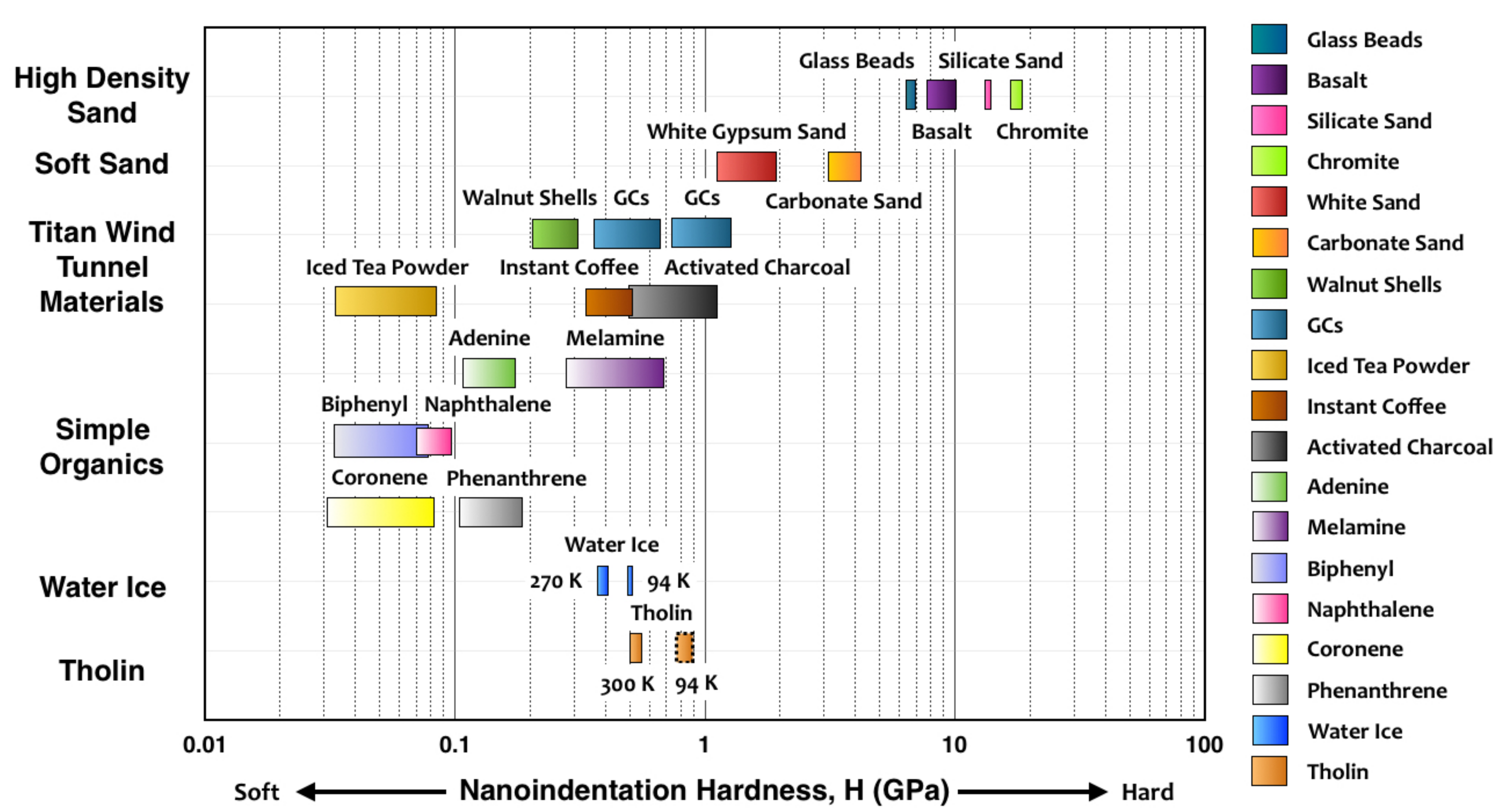}
 \caption{Nanoindentation hardness plot for all tested materials. The color bar includes the standard deviations from measurements for each material. GCs have two characteristic hardness values probably because each GC is a mixture of two substances. Other materials are named and grouped in the same way as Figure \ref{fig:modulus}. For water ice and tholin at 94 K, their hardness values are predicted by using the nanoindentation hardness--modulus relationship in Figure \ref{fig:relationship}.}
 \label{fig:hardness}
 \end{figure}

The elastic moduli and nanoindentation hardnesses of all of the materials are shown in Figure \ref{fig:modulus} and Figure \ref{fig:hardness}. Tholin film has a Young's modulus of 10.4$\mathrm{\pm}$0.5 GPa and hardness of 0.53$\mathrm{\pm}$0.03 GPa, and tholin particles have similar values. A comparison of the load-displacement curves for tholin and fused silica (modulus 72.3$\mathrm{\pm}$0.2 GPa, hardness 9.5$\mathrm{\pm}$0.1 GPa) is shown in Figure \ref{fig:load_displacement_curve}(b). Tholin has smaller maximum indentation load, smaller stiffness, and larger contact area compared to fused silica, which results in smaller hardness and elastic modulus values. However, amorphous organics/polymers (tholin is an amorphous solid, Quirico et al., 2008) usually have moduli in the range of $\mathrm{10}^{-3}$--10 GPa (Meyers and Chawla, 2009), tholin's elastic modulus is on the high end, indicating its large stiffness among this type of material. This may be caused by cross-linking between molecule chains in tholin similar to network polymers (Dimitrov \& Bar-Nun, 2002). The high density of cross-linking makes sliding of molecules difficult, so stretching or breaking of covalent bonds is necessary to deform tholin.

Even though tholin is very stiff as an organic material, its elastic modulus and hardness are an order of magnitude lower than silicate beach sand (modulus $\sim$100 GPa, hardness $\sim$14 GPa) and basalt (modulus $\sim$100 GPa, hardness $\sim$9 GPa). As a mechanically weak sand on Earth, white gypsum is an example of a material that is not able to transport long distances because of its mechanical weakness and also its high solubility in water (Lorenz \& Zimbelman, 2014). However, white gypsum sand has larger stiffness (37 GPa) and hardness (1.5 GPa) than tholin, as is also true for carbonate sand (modulus $\sim$74 GPa, hardness $\sim$3.7 GPa).

It is interesting to note that lots of low density wind tunnel materials have a similar elastic modulus and hardness to tholin, including walnut shells (modulus 7 GPa, hardness 0.3 GPa), GCs (modulus 16 or 9 GPa, hardness 1 or 0.5 GPa), instant coffee (modulus 8 GPa, hardness 0.4 GPa), and activated charcoal (modulus 9 GPa, hardness 0.8 GPa). Even though those materials have very different interparticle forces compared to tholin (Yu et al., 2017a, b). In contrast, the high density wind tunnel materials, quartz sand, chromite, and glass beads, have similar hardness and modulus values to silicate sand.

The PAHs and the polyphenyl we used (naphthalene, biphenyl, coronene, and phenanthrene) all have aromatic rings and are only made of carbon and hydrogen. Their indentation hardness and elastic moduli are all lower than those of tholin. The two nitrogen-containing organics we used (adenine and melamine) have been previously detected in tholin samples (H\"orst et al., 2012; He \& Smith, 2013, 2014a, b). Adenine has smaller elastic modulus (4.3$\mathrm{\pm}$0.7 GPa) and hardness (0.14$\mathrm{\pm}$0.03 GPa), while melamine has similar hardness and elastic modulus values (modulus 9.0$\mathrm{\pm}$2.8 GPa, hardness 0.48$\mathrm{\pm}$0.21 GPa) to tholin. This is probably because there is a larger density of hydrogen bonds in melamine than adenine, which makes the structure stronger (Sakurada \& Keisuke, 1975). In addition, melamine can polymerize with agents like formaldehyde and form the one of the strongest network polymers, melamine resins (Jones \& Ashby, 2011), so the existence of melamine in tholin would support tholin's highly cross-linked structure.

\begin{figure}[h]
\centering
\includegraphics[width=30pc]{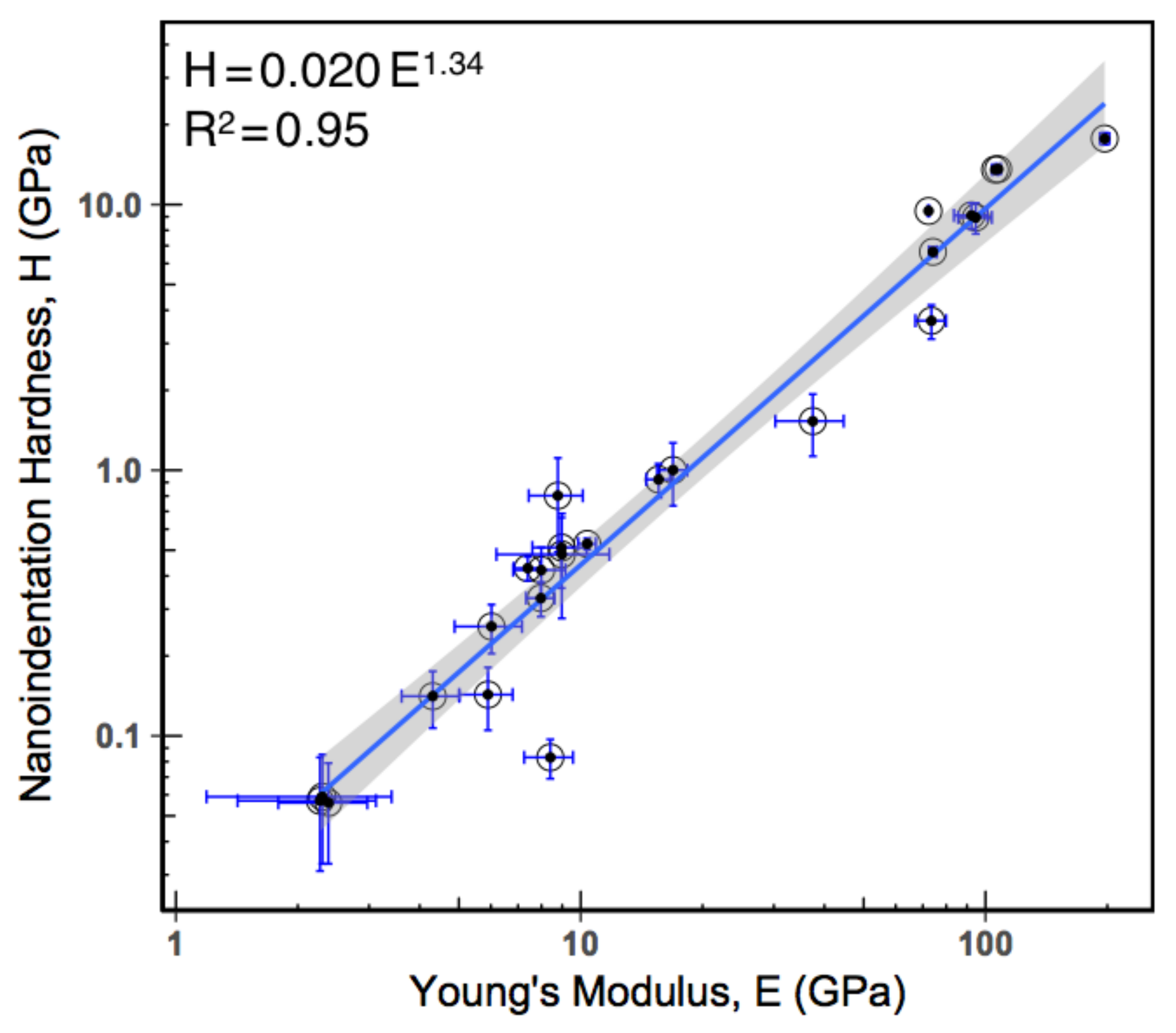}
\caption{Shown here is the logarithmic nanoindentation hardness (H) versus logarithmic elastic modulus (E) values for all test materials and a fitted power law curve (blue line), where $\mathrm{H=0.019E^{1.37},\ R^2=0.95}$. The gray shaded area marks the 95\% confidence intervals for the fitting ([0.012; 0.030] and [1.21; 1.47]).}
\label{fig:relationship}
\end{figure}

Previous work suggests that indentation hardness is correlated with elastic modulus (Labonte et al., 2017). Here we fit our nanoindentation hardness and Young's modulus with a power law equation, $\mathrm{H=0.020E^{1.34},\ R^2=0.95}$, shown in Figure \ref{fig:relationship}. Using this relationship, we can predict the nanoindentation hardness of a material given its elastic modulus value, and vice versa.
 
\subsection{Fracture Toughness}
\begin{figure}[h]
\centering
\includegraphics[width=37pc]{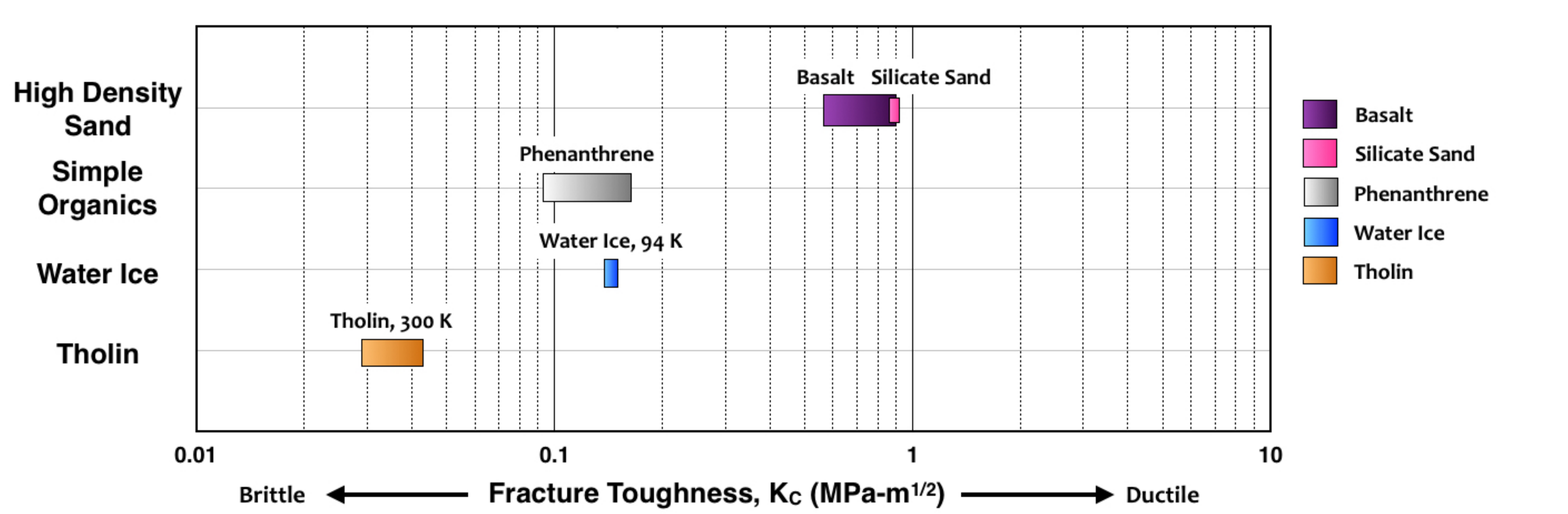}
\caption{Fracture toughness plot for selected materials. The color bar includes the standard deviation values for each material. The selected materials are grouped and named the same as Figure \ref{fig:modulus}. The fracture toughness value of water ice was adopted from Litwin et al. (2012).}
\label{fig:fracture_toughness}
\end{figure}

Since fracture toughness requires a highly smooth surface, we are only able to measure it for several selected materials; the results are shown in Figure \ref{fig:fracture_toughness}. Fracture toughness is an intrinsic material property that describes the resistance of a material to failure. Materials with lower fracture toughness are more brittle. Tholin has a fracture toughness of only 0.036 MPa$\cdot$$\mathrm{m^{1/2}}$, which is much lower than the fracture toughness of typical organic/polymeric materials (0.6--5.0 MPa$\cdot$$\mathrm{m^{1/2}}$). The pop-in events in the loading portion of the load-displacement curve of tholin (Figure \ref{fig:load_displacement_curve}b) also indicate its brittle nature. The fracture toughness of tholin is also a magnitude lower than quartz sand (0.89 MPa$\cdot$$\mathrm{m^{1/2}}$) and basalt (0.55 MPa$\cdot$$\mathrm{m^{1/2}}$). Thus, tholin is much more brittle than silicate sand and is more likely to break apart during transportation. This is consistent with tholin's highly cross-linked structure inferred from the elastic modulus and hardness measurements. High-density cross-linking of molecular chains will provide adequate modulus and strength, but will also lead to extreme brittleness (Meyers and Chawla, 2009). Tholin has a larger elastic modulus and smaller fracture toughness compared to a typical network polymer, epoxy (modulus 2.1--5.5 GPa, fracture toughness 0.3--0.6 MPa$\cdot$$\mathrm{m^{1/2}}$, Meyers and Chawla, 2009), this indicates tholin has much more complex cross-linking compared to regular network polymers. The simple organic we tested, phenanthrene, has a higher fracture toughness than tholin but is still much more brittle compared to silicate sand and basalt. We cannot induce cracks in walnut shell particles even with the highest load (50 mN), probably because they are very porous and ductile, so a much higher load is needed to fracture them. 

\section{Discussion}
\subsection{Temperature's Effect}
Temperature generally has little effect on the mechanical properties of materials when the temperature is lower than a material's phase transition temperature. Elastic modulus and hardness increase slightly with decreasing temperature. For metals and ceramics, the elastic modulus and hardness increase approximately linearly with decreasing temperature from the melting temperature ($\mathrm{T_m}$) (Courtney, 2000). The relationship can be expressed as: 
\begin{linenomath*}
\begin{equation}
E\cong E_0(1-0.5\frac{T}{T_m}),
\label{eq:temperature}
\end{equation}
\end{linenomath*}
where $\mathrm{E_0}$ is the elastic modulus of the material at 0 K. With an elastic modulus value (E) at a given temperature (T), and with the material's melting temperature ($\mathrm{T_m}$), using Equation {\ref{eq:temperature}}, we can calculate $\mathrm{E_0}$ and then estimate elastic modulus at other temperatures. For amorphous polymers, the glassy transition temperature ($\mathrm{T_g}$) is a critical temperature instead of $\mathrm{T_m}$ (Courtney, 2000). Below $\mathrm{T_g}$, polymers are in the glassy regime, have relatively high elastic modulus and hardness and are generally brittle. In this regime, Equation \ref{eq:temperature} holds true for most polymers, with $\mathrm{T_g}$ replacing $\mathrm{T_m}$. While above $\mathrm{T_g}$, the elastic modulus of polymers can decrease by several (3 to 4) orders of magnitude and they become rubbery, this is called the rubbery regime. Here all the experiments we performed were under room temperature ($\sim$300 K) while on Titan the surface temperature is much lower (94 K), so we need to translate our experimental results to Titan conditions. Tholin is a stable solid at room temperature and it does not melt up to at least $\sim$350 K (He \& Smith, 2014c). According to the fracture toughness test, tholin is very brittle and is unlikely to be in its rubbery regime. Here we use the critical temperature $\mathrm{T_g\ or\ T_m=}$350 K for tholin in Equation \ref{eq:temperature}. With the measured modulus value at a temperature of $\sim$300 K, we can estimate the modulus of tholin at 94 K to be around 16 GPa (15.73$\pm$0.79 GPa), shown in Figure \ref{fig:modulus}. Using the fitted linear relationship between the elastic modulus and the nanoindentation hardness in Figure \ref{fig:relationship}, the nanoindentation hardness for tholin at 94 K can be estimated to be around 0.8 GPa (0.83$\pm$0.06 GPa), and is then plotted in Figure \ref{fig:hardness}. The brittleness of glassy polymers would be higher with decreasing temperature; thus tholin should have an even lower fracture toughness at 94 K, which means it would be even more brittle.

\subsection{Candidates for Titan Sand}

There are a few candidates for Titan sand, tholin-like complex organics formed by photochemistry and then modified on the surface, evaporites formed from evaporation process in Titan's dried lake beds, and water ice bed rock. Here we define a material as a good candidate for Titan sand if it is mechanically strong enough to be transported for long distances. We expect the material to have higher hardness and lower brittleness, so it could resist abrasion and impact and will be less likely to be fragmented to dust.

The composition of Titan sand could be similar to tholin if the sand grains are formed by sintering, lithification and erosion, or flocculation (Barnes et al., 2015; Yu et al., 2017a). Flocculation needs liquid methane or ethane to facilitate sand formation. If sand came from, or is coming from, the polar regions of Titan, this mechanism needs the sand to be mechanically strong enough to transport from the polar lakes and seas to the equatorial area, where the dunes are observed. However, our measurements show that tholin is a relatively soft and brittle material compared to common silicate sand and even soft sand like white gypsum sand on Earth. Thus, if Titan sand is similar to tholin, it may not be strong enough to be transported from Titan's poles to the equator, which suggests that Titan sand should be derived close to the equatorial regions. Titan sand could still be formed by the flocculation mechanism if it originated from the tropical lakes (e.g. Griffith et al., 2012) rather than the polar lakes.

Soluble components of aerosols in methane and ethane lakes may form evaporites. The proposed evaporite fields include ancient lake beds in Tui Regio or Hotei Regio (Barnes et al., 2005, 2006; MacKenzie et al., 2014), dried lake beds south of Ligeia Mare (Barnes et al., 2011), and a ring of dried lake bed surrounding Ontario Lacus (Barnes et al., 2009). These proposed evaporite fields are all spectrally bright at 5 $\mu$m. Evaporites are also possible candidates for Titan sand, but they must be physically or chemically modified to become spectrally dark to fit the dune dark spectra. There is no data for mechanical properties of possible Titan evaporites in the solid state, but the simple organic materials we measured have lower elastic moduli and hardnesses than tholin. This suggests that, for evaporites, if they are simple organics as well, may not be strong enough to be transported from the polar regions (e.g. dried lake beds of Ligeia Mare or the ring of dried lake bed around Ontario Lacus) to form the equatorial dunes. This does not rule out the possibility that the evaporites could be transported to the equatorial region from the ancient lake beds that are located close to the equator, such as Tui Regio or Hotei Regio..
 
Another possible candidate for Titan sand is water ice. As Barnes et al., (2008) concluded from VIMS data, water ice cannot be ruled out as a component of Titan sand, since the dark organic sand on Titan could be a result of homogeneously organic-coated water ice grains. On the surface of Titan, water ice is in a hexagonal phase, also known as ice Ih. Proctor (1966) reported measurements of the elastic modulus of monocrystalline ice Ih over a broad temperature range from 40 K to 240 K, with the elastic modulus of ice gradually increasing with decreasing temperature. Using the elastic constants measured by Proctor (1966), we can estimate the elastic modulus of water ice at Titan's surface temperature (94 K) using the method described by Anderson (1963), which is around 11 GPa, shown in Figure \ref{fig:modulus}. From the linear correlation of elastic modulus and hardness in Figure \ref{fig:relationship}, we can estimate the nanoindentation hardness of water ice under Titan's low temperature to be around 0.5 GPa, shown in Figure \ref{fig:hardness}. We can also estimate the elastic modulus and hardness of water ice near its freezing point (270 K), which are surprising only slightly lower than at 94 K (E$\sim$9 GPa and H$\sim$0.4 GPa), shown in Figure \ref{fig:modulus} and \ref{fig:hardness}, as well. The fracture toughness of water ice is nearly invariant with changing temperature and is around 0.15 MPa$\cdot$$\mathrm{m^{1/2}}$ (Litwin et al., 2012). Water ice has a lower elastic modulus and hardness than tholin, but tholin is more brittle. Thus we cannot interpret which material is a better candidate for Titan sand using only their mechanical properties.

Tholin, the simple organics, and water ice on Titan are all mechanically weak, and they may be unable to be transported long distances on Titan. Thus, formation of the materials through subaqueous mechanisms are not favorable for explaining the equatorial dunes on Titan. Several past studies point out that there is no evidence of sediment transportation from Titan's polar regions all the way to the equator (Charnay et al., 2015; Malaska et al., 2016). Solomonidou et al. (2018) and Brossier et al., (2018) suggest that the source of the dune-forming materials may be close to the equatorial region rather than the higher latitudes. This work supports those studies based on the mechanical weakness of Titan sand candidates. It also indicates that sand on Titan maybe produced near where it is observed.

Our study also provides insight into the transportation capacity of Titan sand in the equatorial area. Barnes et al. (2015) proposed that Titan's sands could be in a global transportation system where sand particles should be able to move thousands of kilometers west to east across the equatorial area. However, our measurements suggest that the `fresh' Titan sand may not be strong enough to be transported the long distances as suggested by Barnes et al., (2015). Alternatively, the sand on Titan could be ancient and was chemically/physically modified to be stronger than the `fresh' sand. 

\section{Conclusion}
To understand the origin of Titan's sand, we used nanoindentation to study the mechanical properties of several Titan sand analogs, including tholin thin films, a few organics (PAHs, polyphenyls, and nitrogen containing organics), natural sands on Earth (silicate beach sand, carbonate sand, and white gypsum sand), and some common materials used in the Titan Wind Tunnel (such as walnut shells). Mechanical properties measured include elastic modulus (E, elastic property), hardness (H, plastic property), and fracture toughness ($\mathrm{K_c}$, brittleness, fracture property). Under room temperature, tholin has an elastic modulus of around 10 GPa, nanoindentation hardness of around 0.5 GPa, and fracture toughness of around 0.036 MPa$\cdot$$\mathrm{m^{1/2}}$. Extrapolated to Titan conditions (94 K), tholin's elastic modulus is around 16 GPa, nanoindentation hardness is around 0.8 GPa, and its fracture toughness will be lower than 0.036 MPa$\cdot$$\mathrm{m^{1/2}}$. Compared to common polymers, tholin is very stiff, strong but brittle, which indicates it has much more complex cross-link networks than common network polymers like epoxy resin. Many low density materials used in the Titan Wind Tunnel, such as walnut shells, GCs, instant coffee, and activated charcoal, have similar elastic modulus and hardness values to tholin, which suggests that they are good analogs to Titan sand in terms of their mechanical properties, although their interparticle forces are very different (Yu et al., 2017a). 

We define a material to be a good candidate for Titan sand if it is mechanically strong enough (with high hardness and low brittleness) to be transported for long distances without being abraded to dust. However, the elastic modulus and hardness values of natural sand on Earth are an order of magnitude larger than tholin: silicate beach sand has an elastic modulus of over 100 GPa and hardness of around 10 GPa; even the mechanically weak white gypsum sand has a higher elastic modulus and hardness than tholin (E=37 GPa and H=1.5 GPa). Tholin is also much more brittle than silicate sand: its fracture toughness is an order of magnitude smaller than silicate sand ($\mathrm{K_c=0.9\ MPa\cdot m^{1/2}}$). This indicates that the organic sand (if it is compositionally similar to tholin) on Titan may not originate from the current lakes and seas on Titan; being soft and brittle, it is not mechanically strong enough to transport from the pole to the equator. The elastic moduli and hardness of the simple organics are all lower than tholin, which indicates evaporites (if they are made of simple organics), may not be good candidates for Titan sand, unless they are physically/chemically modified to be more complex and mechanically stronger. Water ice has similar elastic modulus and hardness values to tholin (E=11 GPa, H=0.5 GPa), but it has a slightly higher fracture toughness ($\mathrm{K_c=0.15\ MPa\cdot m^{1/2}}$). However, we are unable to determine whether water ice or tholin is a better candidate for Titan sand by only comparing their mechanical properties.

\acknowledgments
Data of the measured nanoindentation hardness, elastic modulus, and fracture toughness values of materials are available in the paper and in the supporting information. The original indentation data and crack images (for fracture toughness calculations) can be found the repository with the following DOI: https://dx.doi.org/10.7281/T1/TP9B4Y. Partial support to X. Yu is provided by NASA grant NNX14AR23G. C. He is supported by the Morton K. and Jane Blaustein Foundation. We thank Maya Gomes for providing the carbonate sand from Bermuda and Xu Yang for providing technical support with R programming. We continue to be thankful beyond words for Nathan Bridges, who although did not live to see this work, started us on the journey that led us here.

\end{document}